\documentclass[preprint,12pt]{elsarticle}
\usepackage{graphicx}
\usepackage{amssymb}

\journal{Physica A}

\begin{document}

\begin{frontmatter}

\title{Bounds of percolation thresholds in the enhanced binary tree}

\author{Seung Ki Baek\corref{cor1}}
\ead{garuda@tp.umu.se}
\author{Petter Minnhagen}
\cortext[cor1]{Corresponding author}

\address{Integrated Science Laboratory, Department of Physics, Ume{\aa} University, 901 87 Ume{\aa}, Sweden}

\begin{abstract}
By studying its subgraphs, it is argued that
the lower critical percolation threshold of the enhanced binary tree (EBT)
is bounded as $p_{c1} < 0.355059$, while
the upper threshold is bounded both from above and below by $1/2$
according to renormalization-group arguments. We also review a
correlation analysis in an earlier work, which claimed a
significantly higher estimate of $p_{c2}$ than $1/2$, to show that this
analysis in fact gives a consistent result with this bound. Our result
confirms that the duality relation between critical thresholds does not hold
exactly for the EBT and its dual, possibly due to the lack of transitivity.
\end{abstract}


\end{frontmatter}

\section{Introduction}
\label{sec:intro}

\begin{figure}
\begin{center}
\includegraphics[width=0.23\textwidth]{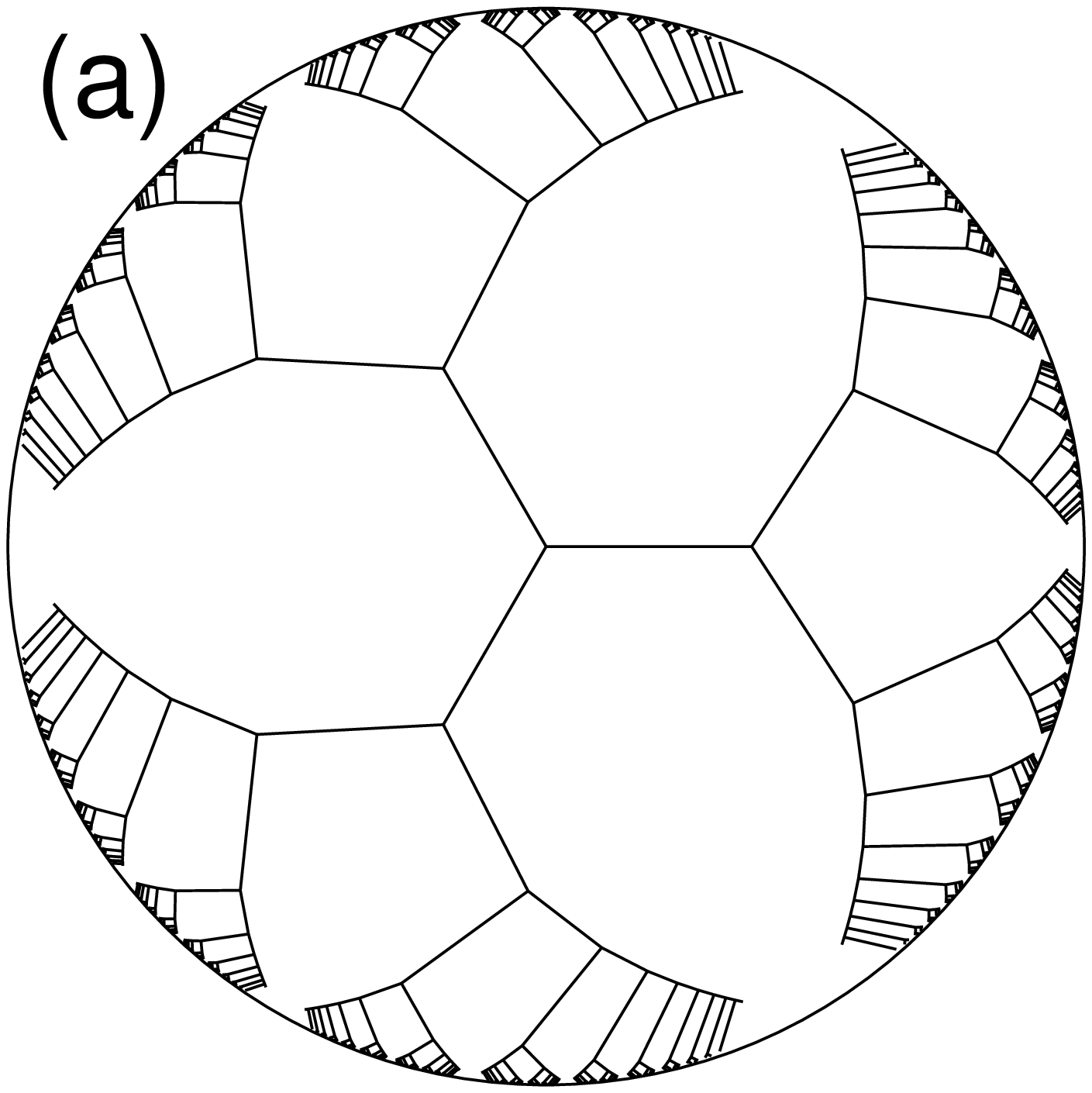}
\includegraphics[width=0.23\textwidth]{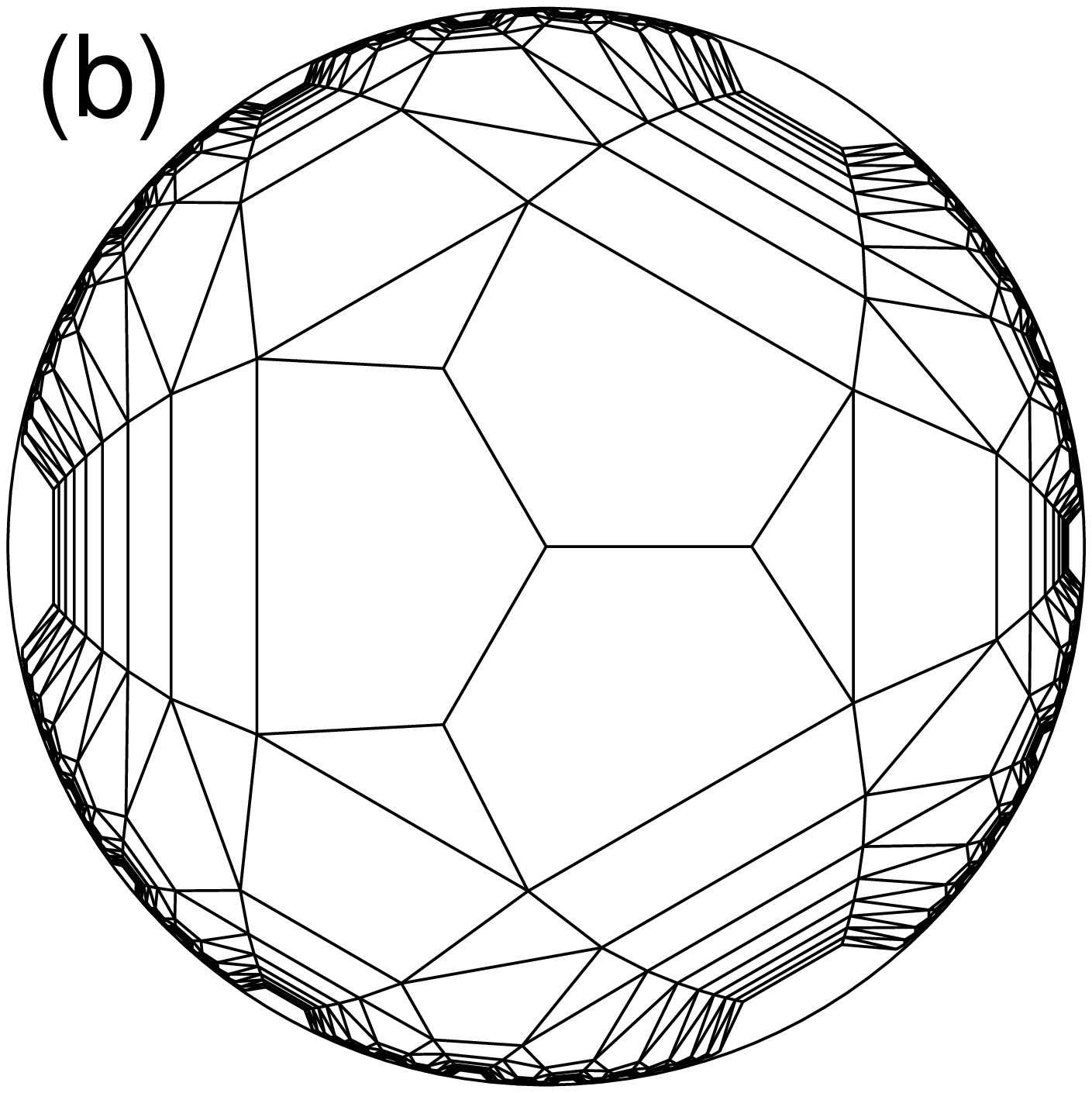}
\end{center}
\caption{Schematic plots of hyperbolic structures drawn on the Poincar\'{e}
disk. (a) A tree structure up to level $L=10$ and (b) the enhanced tree
obtained from (a). The midpoint is located at the origin in each plot.}
\label{fig:poin}
\end{figure}

Percolation has been always of physical interest since it was introduced
as a description for a fluid in random media~\cite{broad}.
The percolation problem has served as one of the basic models in
understanding critical phenomena from a geometric
viewpoint~\cite{stauffer}.
An interesting aspect of the percolation phenomenon is that
a lattice with a constant negative curvature may have two different
thresholds, which coalesce into one as the curvature vanishes~\cite{baek}.
Figure~\ref{fig:poin}(a) gives the simplest example of a hyperbolic lattice:
a regular tree with level $L$, where $L$ is defined as the maximum path
length from the origin. The set of points at the same distance $l$ from the
origin constitutes the $l$th layer.
In this tree, every vertex has three neighbors except at the
boundary, and the number of vertices scales as $N \sim 2^L$.
An immediate consequence is that the number of boundary points, which
scales as $B \sim 2^{L-1}$, always occupies a finite fraction of $N$ even in
the thermodynamic limit. The existence of two percolation thresholds has
been attributed to this peculiar structural property. That is, when the
occupation probability $p$ reaches the lower threshold, $p_{c1}$, a
system-wide connection is first achieved, while the largest cluster
occupies a finite fraction of $N$ only if $p$ exceeds the upper threshold,
$p_{c2}$. Therefore, if we count the boundary points connected to the
midpoint of the system as we increase $p$, the connection simply does not
exist at $p<p_{c1}$, and even after passing this first threshold the
connected boundary points remain as a negligible part of the whole
boundary until $p$ reaches $p_{c2}$.
In the case of the bond-percolation problem in a simple tree as in
Fig.~\ref{fig:poin}(a), for example, one can easily show that
$p^{\rm tree}_{c1} = 1/2$ and $p^{\rm tree}_{c2}=1$~\cite{baek}.

\begin{figure}
\begin{center}
\includegraphics[height=0.08\textheight]{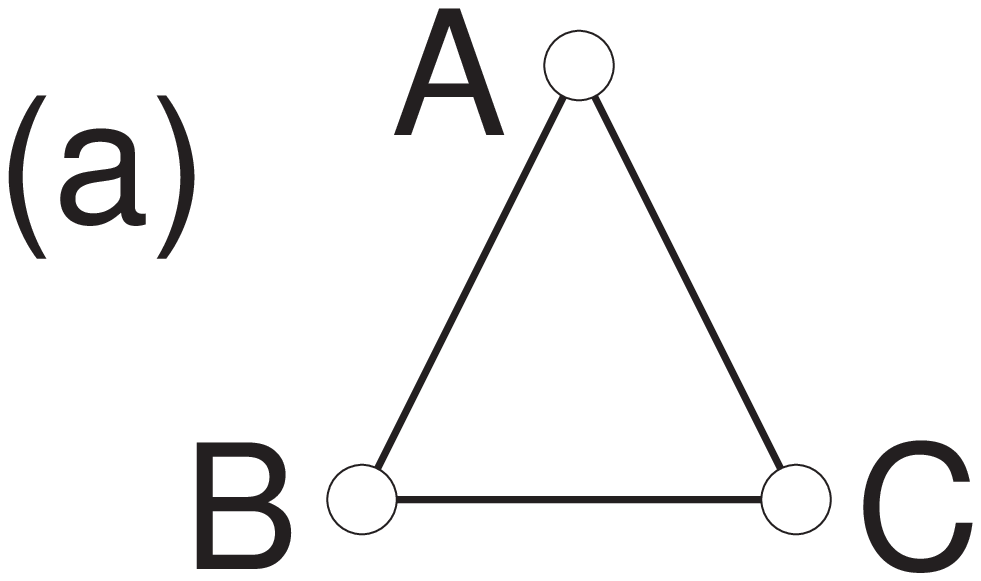}
\includegraphics[height=0.08\textheight]{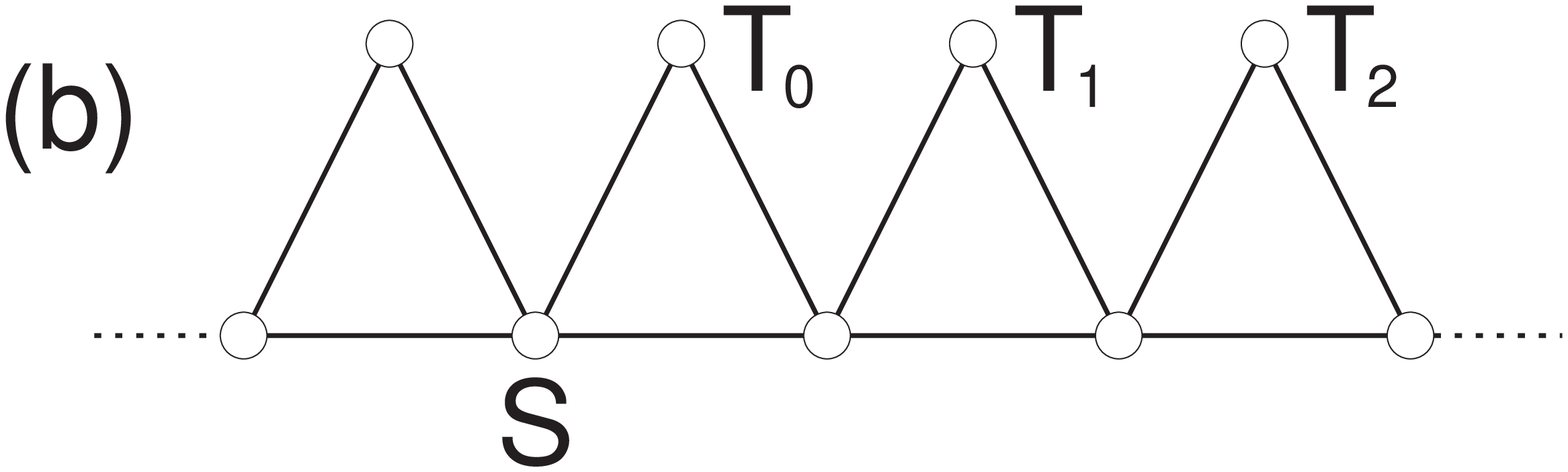}
\end{center}
\caption{(a) A triangular cell having vertices $A$, $B$, and $C$. (b) An
array of such triangular cells, where the $T_i$ indicate the points in
the upper layer on the right-hand side of $S$.}
\label{fig:tr}
\end{figure}

The enhanced binary tree (EBT) is a nontrivial model derived from a
tree~\cite{nogawa}: it is obtained by adding bonds to the tree between every
pair of neighboring points within each layer $l > 1$
[Fig.~\ref{fig:poin}(b)], so it has $p_{c1} < p^{\rm tree}_{c1}$ and $p_{c2}
< p^{\rm tree}_{c2}$ due to the existence of loops.
An interesting analogy of an EBT would be a biological taxonomy which is
mostly a tree structure but with genes horizontally transferred as well
(see, e.g., \cite{gilbert}).
Although the lower
threshold could be easily measured as $p_{c1} = 0.304(1)$~\cite{nogawa},
there remains a controversy in locating $p_{c2}$~\cite{nogawa,com,reply}:
\cite{nogawa} claimed that $p_{c2} = 0.564(1)$, and the correlation
analysis was suggested to support this claim~\cite{reply}, while other
numerical methods preferred $p_{c2} = 0.48(1)$~\cite{com}. Recently, there
appeared an analytic calculation~\cite{future},
which suggests a possible generalization of the triangle-triangle
transformation~\cite{scu,ziff06,ziff09}:
in order to use this transformation in an exact fashion,
one should be able to decompose a given structure of
identical triangular unit cells, where the three vertices of one unit cell
are denoted as $A$, $B$, and $C$, respectively (see Fig.~\ref{fig:tr}(a)).
Then $P(A,B,C)$ is defined as the probability that
$A$, $B$, and $C$ are all connected,
$P(\bar{A}, B, C)$ as
the probability that $B$ and $C$ are connected but $A$ is disconnected from
them, and $P(\bar{A}, \bar{B}, \bar{C})$ as
the probability that none of them are connected to each other.
One may locate the critical
threshold by equating these two probabilities~\cite{ziff06}:
\begin{equation}
P(A,B,C) = P(\bar{A}, \bar{B}, \bar{C}).
\label{eq:ziff}
\end{equation}
From a trivial equality, $P(\bar{A}, \bar{B}, \bar{C}) + P(A,B,\bar{C}) +
P(A,\bar{B},C) + P(\bar{A},B,C) + P(A,B,C) = 1$,
one can express Eq.~(\ref{eq:ziff}) as
\begin{equation}
\frac{P(A,B)}{1-P(\bar{A},B,C)}+ \frac{P(A,C)}{1-P(\bar{A},B,C)}
\equiv \Psi = 1,
\label{eq:psi}
\end{equation}
where $P(\alpha,\beta)$ means the probability of connection
between vertices $\alpha$ and $\beta$.
If one picks up an arbitrary starting point $S$ in a very large layer as in
Fig.~\ref{fig:tr}(b), 
the probability of connecting to the upper layer
on its right-hand side can be written as $\sum_{i=0}^{\infty} \mbox{Prob}(T_i$
is the first right-hand connection$) = \sum_{i=0}^{\infty}
P(A,B) P^i(\bar{A},B,C)$, which is identical to the first term in
Eq.~(\ref{eq:psi}). Likewise, the second term in Eq.~(\ref{eq:psi}) means
the probability of a left-hand connection from $S$. In this sense,
Eq.~(\ref{eq:ziff}) can be regarded as
describing a certain connective property between two adjacent layers at
criticality (see also \cite{vuo}, where one finds a similar idea).
Applying this idea to the EBT
\emph{without} requiring the self-duality of the triangular type,
one obtains $\Psi(p) = p(1+p)/[1-p(1-p)]$~\cite{future}.
Then, the equation $\Psi(p)=1$ is satisfied at $p=1/2$, which in
\cite{future} was interpreted as an exact value of $p_{c2}$ for the
EBT. Since the simple binary tree has $\Psi^{\rm
tree}(p^{\rm tree}_{c1})=1/2$, assuming this again holds for the EBT,
one obtains the value of $p_{c1}$ as $(\sqrt{13}-3)/2 \approx
0.302776$~\cite{future}.

In this work, we use alternative approaches and argue that $p_{c1} <
0.355059$ by examining solvable subgraphs and $p_{c2} \le 1/2$ by
means of the renormalization-group (RG) method on hierarchical
structures~\cite{boet}. We also demonstrate that the correlation analysis
indeed gives a consistent result with this argument for $p_{c2}$. The paper
is organized as follows. We explain the results for $p_{c1}$ and $p_{c2}$ in
Sec.~\ref{sec:sub} and Sec.~\ref{sec:rg}, respectively. Then
Sec.~\ref{sec:cor} additionally discusses the correlation
behavior, and Sec.~\ref{sec:sum} summarizes this work.

\section{Lower threshold}
\label{sec:sub}

A critical percolation threshold becomes higher when part of the links
are removed from the original graph. We could therefore
argue that $p_{c1} < 1/2$ above by considering the simple
binary tree as a subgraph of the EBT. We refine this bound by taking
larger but still solvable subgraphs.

\begin{figure}
\begin{center}
\includegraphics[width=0.32\textwidth]{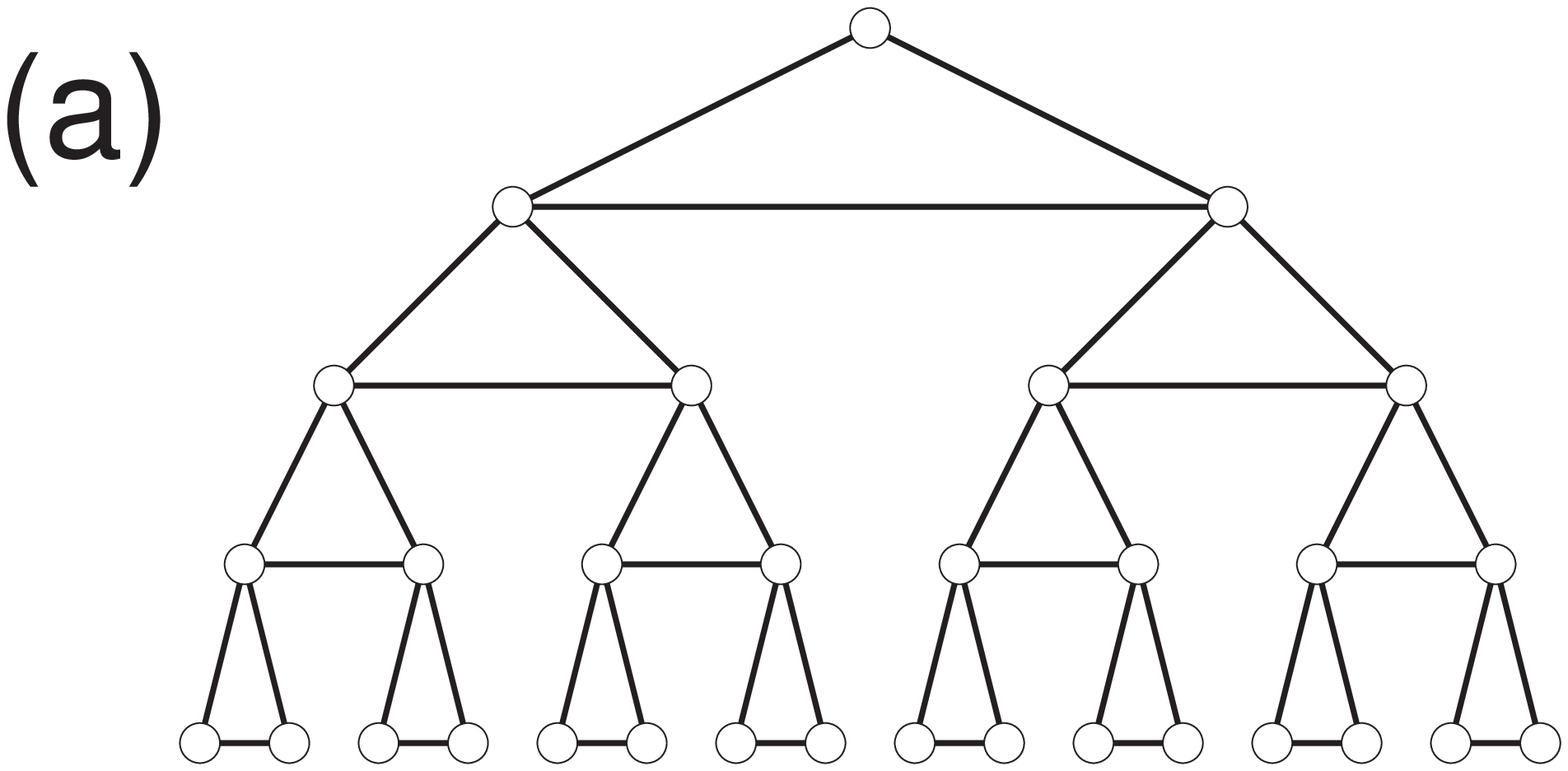}
\includegraphics[width=0.32\textwidth]{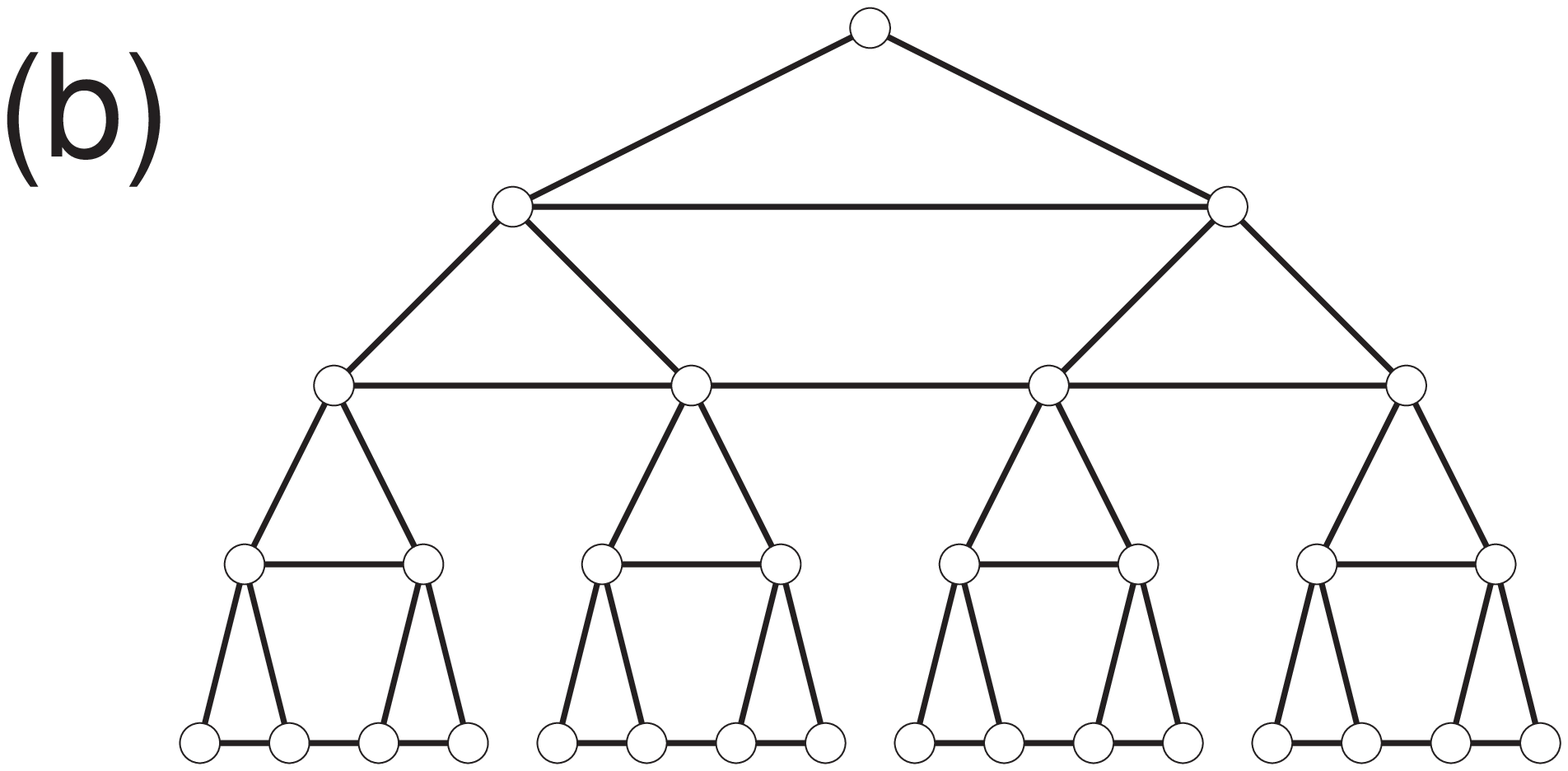}
\includegraphics[width=0.32\textwidth]{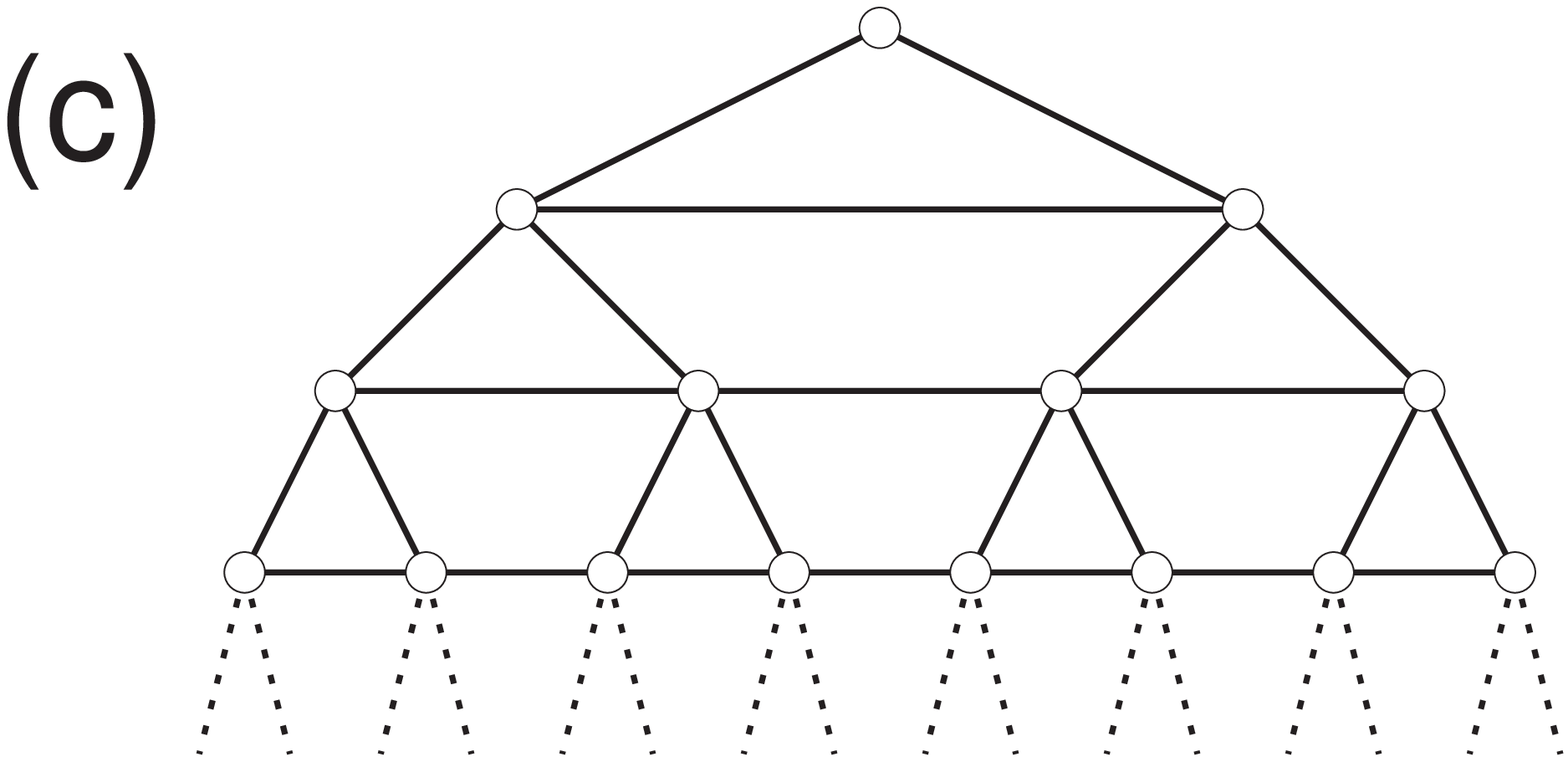}
\end{center}
\caption{Subgraphs of the EBT for bounding $p_{c1}$. (a) The tree made of
triangles has its lower threshold at $p^{\ast}_1 \approx 0.403032$, which
bounds $p_{c1}$ from above. (b) By adding more bonds at every second layer,
a larger unit cell is constructed, and we get a sharper bound, $p^{\ast}_2
\approx 0.373897$. (c) The largest unit cell considered in this work,
giving the upper bound $p_{c1} < p^{\ast}_3 \approx 0.355059$. The dotted
lines mean that other unit cells are attached there.
}
\label{fig:sub}
\end{figure}

One may first consider a tree of triangles, which is also a subgraph of
the EBT (see Fig.~\ref{fig:sub}(a)). We are going
to ask whether there can be any path penetrating from the top to the bottom.
A relevant quantity would then be how many vertices can be
reached on the $(l+1)$th layer from a single vertex on the $l$th layer.
If we focus on a unit cell, which is simply a triangle here,
we find $2^3$ possible configurations, since it has three bonds.
By checking all of these configurations, one can easily get the average
number of descendants, i.e., the expected number of bottom points connected
to the top vertex within the cell. It is easily given as
\[n_1(p) = 2p(1-p)^2 + 6p^2 (1-p) + 2p^3 = 2p \left[1+p(1-p) \right].\]
Solving $n_1(p^{\ast}_1)=1$, we obtain $p^{\ast}_1  = \frac{1}{3} \left(
1 + 2\cos\theta - 2\sqrt{3} \sin\theta \right) \approx 0.403032$, where
$\theta \equiv \frac{1}{3} \arctan \left( \frac{9 \sqrt{37}}{5\sqrt{3}}
\right)$. This provides an improved upper bound for
$p_{c1}$. Note also that $n_1(p)$ is a monotonically increasing function of
$p$ from zero to 2. This implies that the upper threshold of this subgraph
is located at $p=1$, since otherwise the fraction of connected vertices will
decrease as the number of passing layers increases.

This sort of construction requires that a unit cell should possess only one top
vertex and that every bottom-layer point equally becomes a new top vertex for
a subsequent unit cell. We can therefore extend this idea a little further,
as follows. For every second layer,
we add a bond between two daughter triangles under the same mother triangle,
as in Fig.~\ref{fig:sub}(b). This creates a new unit cell containing three
triangles plus one bond between the daughter cells.
Such a cell has $10$ bonds in total, meaning
$2^{10}=1024$ possible configurations. Directly enumerating them shows again
how the top vertex connects to the bottom layer within this cell. The
average number of descendants then reads
\[ n_2(p) = 2p^2 \left[ 1+p(1-p) \right] \left[2+ p(1-p)(3 +3p +p^2 -10p^3 +5p^4)\right], \]
which ranges over $[0,2^2]$.
Numerically solving $n_2(p^{\ast}_2)=1$ leads to $p^{\ast}_2 \approx 0.373897$.
It is still possible to consider a larger unit cell with seven triangles
(see Fig.~\ref{fig:sub}(c)) and find the average number of connected
bottom-layer points as
\begin{eqnarray*}
n_3(p) &=& 2p^3 \left[1+p(1-p) \right] \left[ 4 + p(1-p)(13 +26p +23p^2 -37p^3
-64p^4\right.\\
&&-42p^5 -116p^6 +234p^7 +469p^8 -830p^9 +1811p^{10} -2898p^{11}\\
&& -4735p^{12} +21801p^{13} -31538p^{14} +24399p^{15} -10894p^{16}\\
&&\left.+2664p^{17} -278p^{18} )  \right]
\end{eqnarray*}
from $2^{25}$ configurations.
By solving $n_3(p^{\ast}_3)=1$, we find an upper bound for $p_{c1}$ as
$p^{\ast}_3 \approx 0.355059$. On the other hand, it is readily seen that
$p_{c1}$ can never be lower than $0.25$, since every vertex in the EBT has
$z=5$ neighbors except the zeroth layer, so that the tree approximation
yields $1/(z-1) = 1/4$. We note that both the numerically obtained threshold
$p_{c1} = 0.304(1)$~\cite{nogawa} and the analytic prediction $p_{c1} =
(\sqrt{13}-3)/2 \approx 0.302776$~\cite{future} lie within these upper and
lower bounds. Since this subgraph method soon becomes impractical due to the
huge number of possible configurations as the size of a unit cell grows,
there is no sharper bound available at present. One may try an
extrapolation by using the correlation-length scaling exponent $\nu=1$ at
$p=p_{c1}$~\cite{baek}, but this will be a rather crude estimation.
It is also clear that
these subgraphs do not delimit $p_{c2}$ of the EBT at all since all of
them have their upper thresholds at $p=1$.

\section{Upper threshold}
\label{sec:rg}

\begin{figure}
\begin{center}
\includegraphics[width=0.33\textwidth]{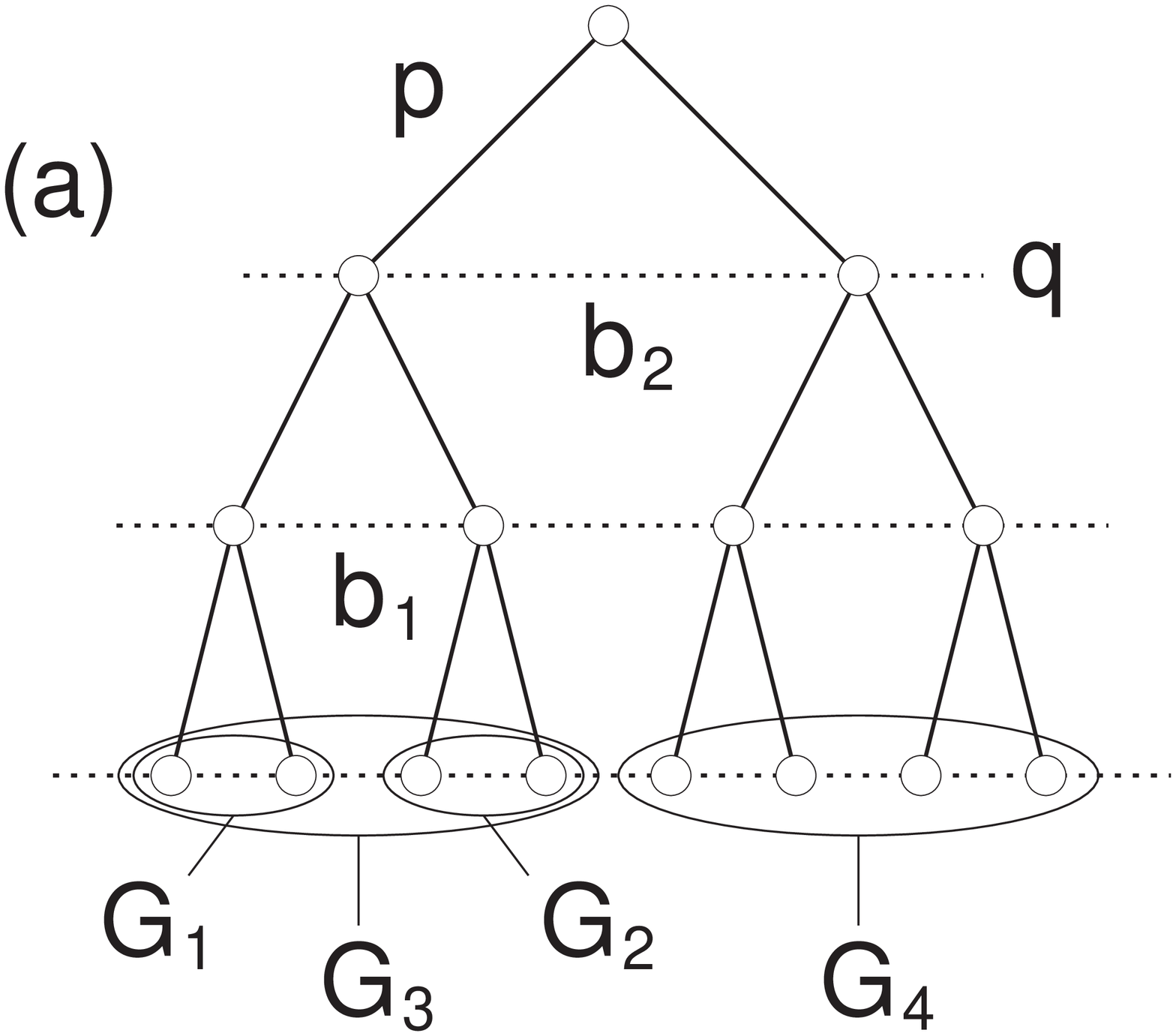}
\includegraphics[width=0.27\textwidth]{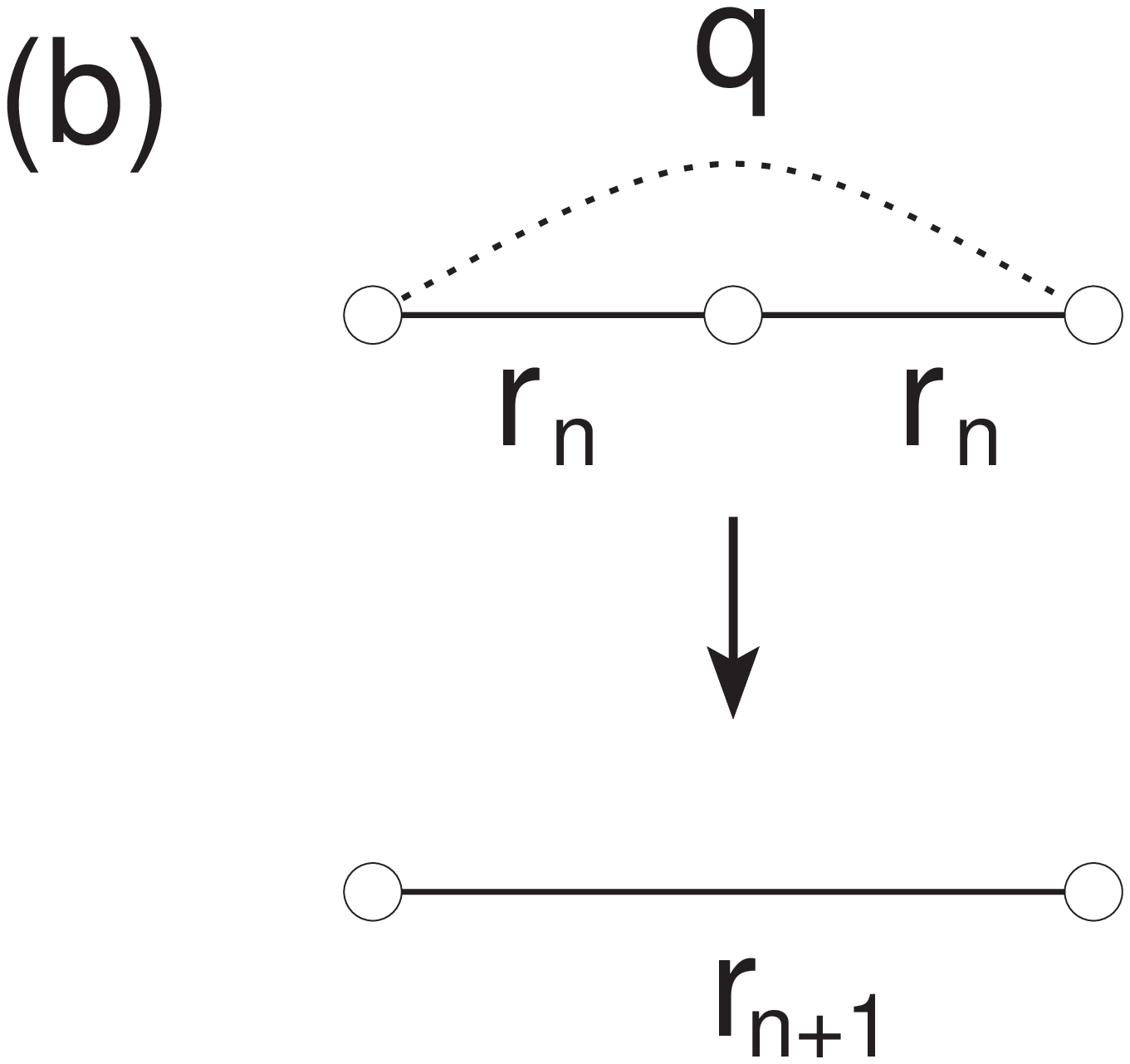}
\includegraphics[width=0.34\textwidth]{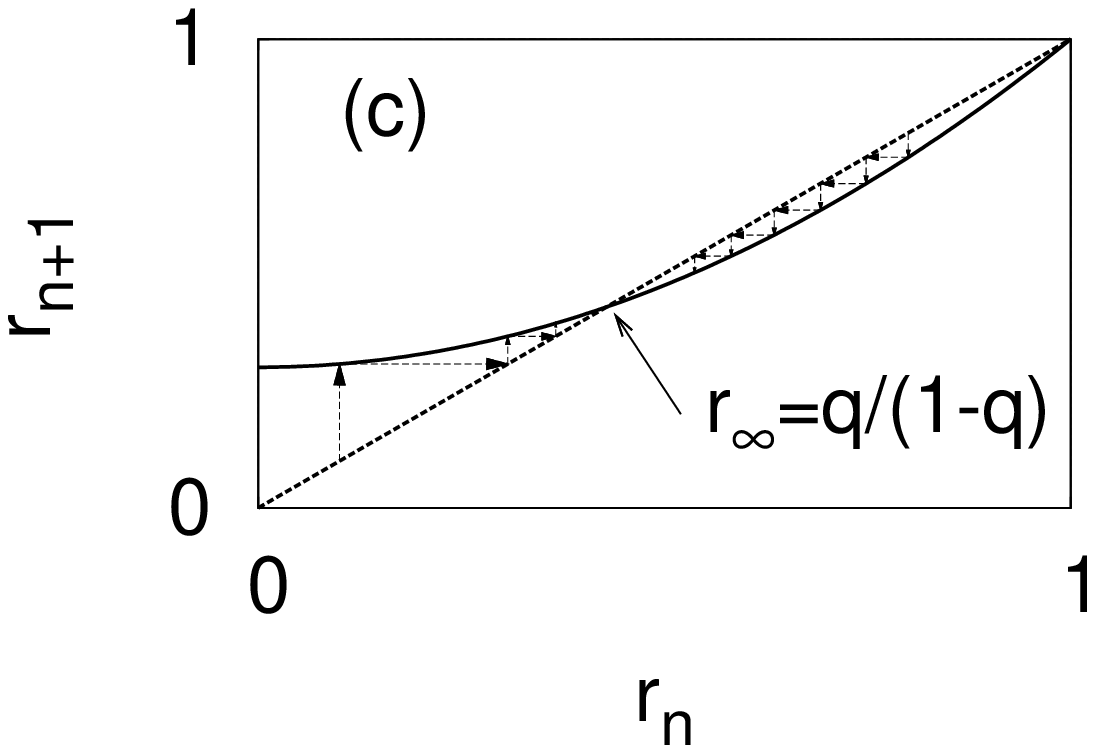}
\includegraphics[width=0.27\textwidth]{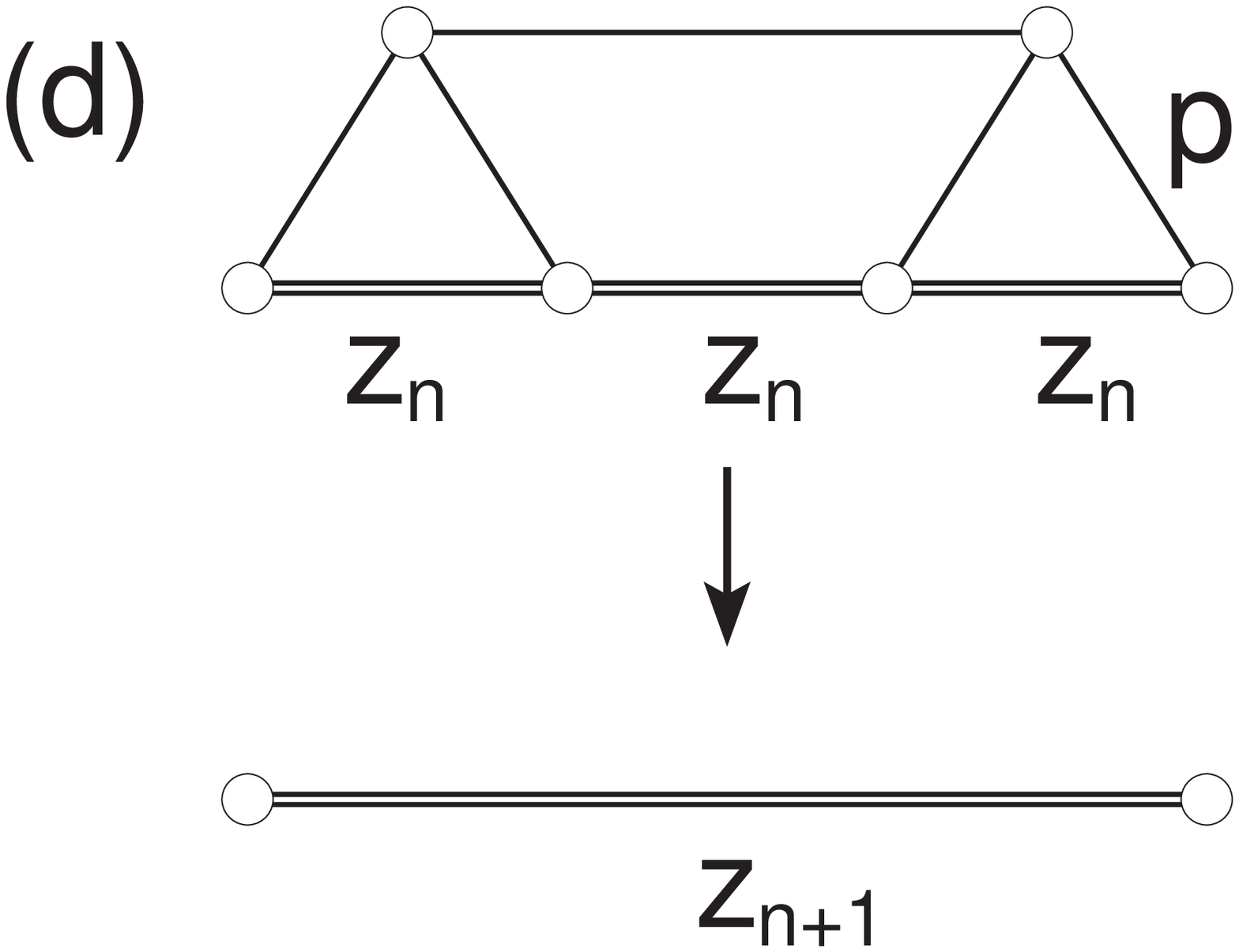}
\end{center}
\caption{(a) A part of an EBT structure. The
tree part (solid) is occupied with probability $p$ and the rest of the bonds,
i.e., the horizontal lines, are occupied with probability $q$. At $p=1/2$,
a single bond $b_1$ will directly connect two groups of boundary points,
$G_1$ and $G_2$, with probability $q$, while $b_2$ will connect $G_3$ and
$G_4$.
(b) Schematic description for $r_{n+1}$, the global-connection probability
for the bottom layer at the $(n+1)$th RG iteration. Hence the probability
$r_n$ at the $n$th iteration is related to $r_{n+1}$ by Eq.~(\ref{eq:rg}).
(c) Dynamics of the RG flow for $q \le 1/2$ according to Eq.~(\ref{eq:rg}),
which is here drawn as the solid line.
(d) An alternative recursion scheme illustrating Eq.~(\ref{eq:rg2}), where
the double lines represent coarse-grained links, i.e., $z_n$ or $z_{n+1}$.
}
\label{fig:latt}
\end{figure}

The main idea in bounding $p_{c2}$ begins with the fact that a simple binary
tree has a lower
threshold $p^{\rm tree}_{c1} = 1/2$. We assign occupation probabilities $p$
and $q$ to the tree part and the rest of the EBT, i.e., the horizontal bonds
in Fig.~\ref{fig:latt}(a), respectively.
Let us start from $q=0$, where the EBT becomes identical to a simple tree.
At $p=1/2$, the tree part reaches its first critical threshold so that
each vertex is connected to one boundary point on average. Then we increase
$q$ so that a finite fraction of boundary points can merge into a single
cluster.

An interesting property of the EBT is that the horizontal bonds would have
different connection ranges depending on which layer they belong to.
In Fig.~\ref{fig:latt}(a), bond $b_1$ directly connects
two groups of boundary points, $G_1$ and $G_2$, at average distance 2,
which is measured along the bottom layer. On the other hand, the average
range of connection between groups $G_3$ and $G_4$ by bond $b_2$ is twice
as long as than that. In other words, if we focus only on the
bottom layer by setting $p=1/2$, the horizontal bonds constitute a
hierarchical structure so that the range of connections becomes twice as
long every time when a new higher level is introduced.
Such a property allows one to
formulate an RG equation~\cite{boet}, which here can be written as
\begin{equation}
r_{n+1} = q + (1-q) r_n^2,
\label{eq:rg}
\end{equation}
where $r_n$ is the probability of global connection in the bottom layer at
the $n$th RG scaling transformation (see Fig.~\ref{fig:latt}(b)).
Here, we are asking ourselves how probable it is for two different
points that we have arbitrarily picked up from the boundary bottom layer to
belong to the same cluster. Since a cluster containing the middle part of
the system occupies only a negligible fraction of the whole boundary at
$p<p_{c2}$ by definition, this means that such a chance gets significantly
large above $p_{c2}$, and thereby Eq.~(\ref{eq:rg}) is related to the upper
threshold.
It is stated in Eq.~(\ref{eq:rg}) that the global connection can be
established either by a new long link with probability $q$ or by
existing shorter links.
In the limit of large $n$, we may set $r_n = r_{n+1} = r_{\infty}$ and then
Eq.~(\ref{eq:rg}) is easily solved to yield a nontrivial stable solution,
$r_{\infty} = q/(1-q)$, as shown in Fig.~\ref{fig:latt}(c)~\cite{boet}.
Noting that $r_{\infty}$ is responsible for connecting a number of the
boundary points to one another at criticality, we find that $q=1/2$ should
be a transition point \emph{provided} that $p$ is fixed at $p^{\rm
tree}_{c1} = 1/2$.
Since $p$ and $q$ happen
to have the same value here, even if considering the homogeneous case where
$q$ is always set equal to $p$, we can conclude that $p=1/2$ is high enough
to connect a significant fraction of boundary points. In short, $p=1/2$
should be higher than or equal to $p_{c2}$, the upper critical percolation
threshold of the EBT.

We note that Eq.~(\ref{eq:rg}) is an approximate description with a
coarse-grained variable $r_n$, chosen for ease of explanation. It is
possible to make an alternative recursion scheme which instead yields a
lower bound. The scheme is illustrated in Fig.~\ref{fig:latt}(d), where
$q$ is assumed to be equal to $p$.
There are two outer points on the left-hand
side and two others on the right-hand side. In merging a set of bonds
into a single one with connection probability $z_{n+1}$, we
are interested in linking any of the left outer points to any of
the right outer points. Obviously, a part of contribution to $z_{n+1}$
comes from filling the upper bond with probability $p$. Even if it is not
filled with $1-p$, there remain a few more possible cases.
Suppose that there happen to be two filled bonds which are the right one
of the left triangle and the left one of the right triangle, for example.
It is then enough to have only one $z_n$ in between. Likewise, we can
consider all the other cases and arrive at the following recursion relation:
\begin{equation}
z_{n+1} = p + (1-p) \left[ (1-p)^2 z_n^3 + 2p(1-p) z_n^2 +
p^2 z_n \right].
\label{eq:rg2}
\end{equation}
Solving this by setting $z_n = z_{n+1} = z_{\infty}$ as above, one finds
that
\[ z_{\infty} = \frac{\sqrt{1+4p-6p^2+p^4}}{2(1-p)^2} - \frac{1+p}{2(1-p)}. \]
Again, we see that $z_{\infty} = 1$ at $p \ge 1/2$. However, since the
renormalization includes connections between the lower outer points even if
there is no interlayer connection, the result yields a lower bound of the
threshold.

\section{Correlation Analysis}
\label{sec:cor}

\begin{figure}
\begin{center}
\includegraphics[width=0.45\textwidth]{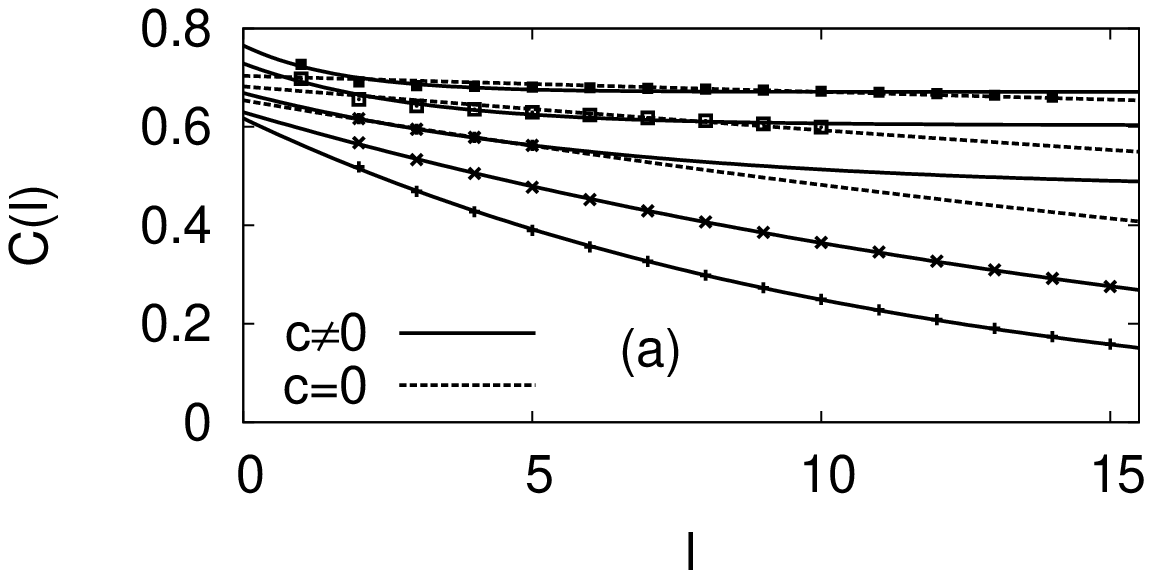}
\includegraphics[width=0.45\textwidth]{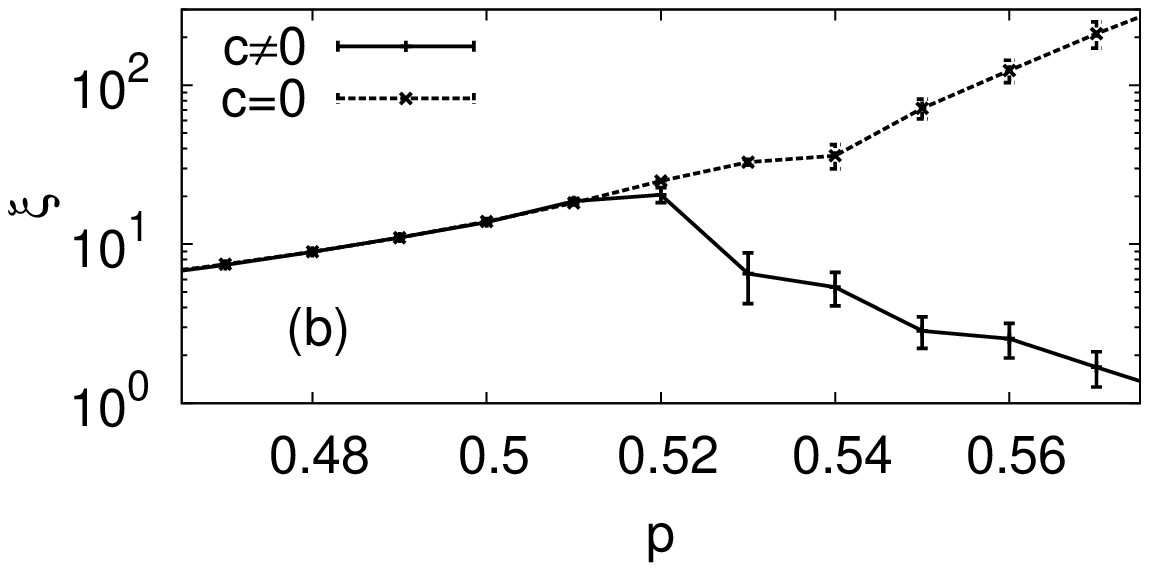}
\includegraphics[width=0.45\textwidth]{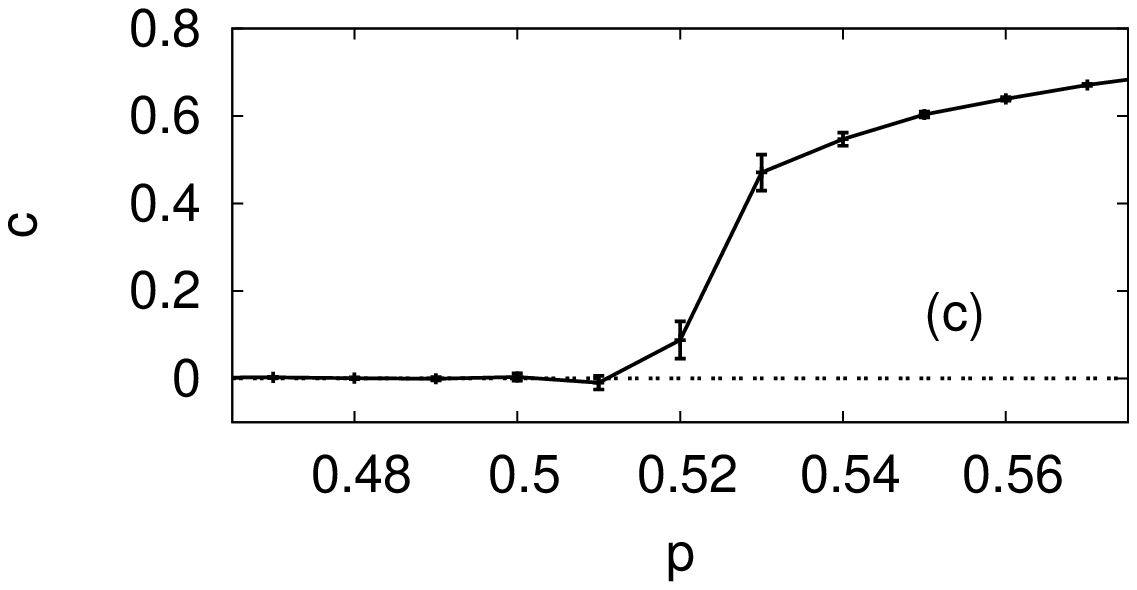}
\includegraphics[width=0.45\textwidth]{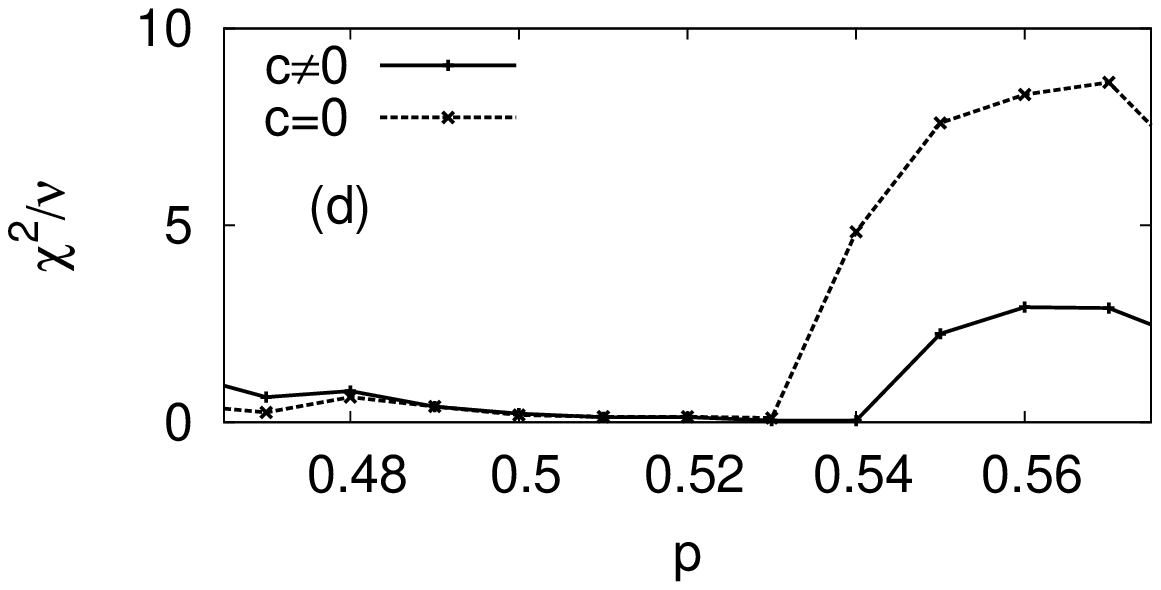}
\end{center}
\caption{ (a) Correlation function measured for $L=24$ over $10^4$ samples.
The occupation probabilities are $p=0.49, 0.51, 0.53, 0.55$, and $0.57$, from
bottom to top.
The solid curves are according to Eq.~(\ref{eq:co}) with least-square-fitted
parameter values, while the dotted curves are found when we fit the data
without $c$. (b) Correlation length $\xi$ obtained from the measurements.
One observes a peak around $p=0.51$
from the fitting by Eq.~(\ref{eq:co}) (solid). If one fits the
same data after setting $c=0$, $\xi$ can be overestimated in the
supercritical phase (dotted).
(c) Asymptotic correlations obtained by Eq.~(\ref{eq:co}).
(d) Reduced $\chi^2$ values for $c\neq 0$ and $c=0$, respectively.}
\label{fig:co}
\end{figure}

A recent numerical estimation in
\cite{reply} suggested $p_{c2} > 0.56$, which certainly exceeds
$1/2$ given above as an upper bound. We present a brief discussion on this
discrepancy. Following
\cite{reply}, let us consider the probability $C(l)$ that points
at level $l$ in the EBT belong to the same cluster as the midpoint does.
Such a correlation will be a monotonically decreasing function of $l$, and
it converges to a constant if $p
> p_{c2}$. If we are to take only robust behavior insensitive to any
particular $L$, a possible way would be to compare two different system
sizes, say $L=23$ and $24$, and take only data points up to $l = l_{\rm
max}$, where they cease to overlap within error bars. Furthermore, we need to
exclude $l<l_{\rm min}$, since the EBT has a lower number of connections
close to the midpoint, so $C(l)$ may decrease anomalously at small $l$.
For the data presented in Fig.~\ref{fig:co}, we set $l_{\rm
min}=2$ for $p<0.55$ and $l_{\rm min}=1$ for $p \ge 0.55$ because of the
more rapid exponential decay, which possibly implies that $\xi \lesssim 2$.
We assume that $C(l)$ within the range of $l_{\rm min} \le l < l_{\rm max}$
will show a simple decaying behavior with a certain characteristic length
scale $\xi$ as
\begin{equation}
C(l) \sim a \exp(-l/\xi) + c,
\label{eq:co}
\end{equation}
where $a$ and $c$ are independent of $l$. The parameter $c$ can be also
said to be the asymptotic value of correlation at large $l$. Since
this term in the fitting procedure was not included in \cite{reply}, 
$\xi$ is presumably overestimated in the supercritical phase, since the
nonvanishing part of $C(l)$ would be interpreted as a very slow decrease
(see Fig.~\ref{fig:co}(a)). Our fitting results plotted in
Fig.~\ref{fig:co}(b) and Fig.~\ref{fig:co}(c) show signatures of
the transition around $p=0.51$, which is significantly lower than $p=0.56$
claimed in \cite{reply}, and is in fact fairly close to $1/2$.
In order to compare the goodness of fits, we calculate the $\chi^2$
statistic defined as
\[
\chi^2 = \sum_i \frac{(O_i - E_i)^2}{\sigma_i^2},
\]
where $O_i$ is the $i$th data point observed with variance
$\sigma_i^2$, and $E_i$ is the corresponding expected value from
Eq.~(\ref{eq:co}).
The number of degrees of freedom, $\nu$, is here given as the number
of data points minus the number of fitting parameters. The reduced $\chi^2$
statistic, $\chi^2/\nu$, allows one to compare the performance of different
fitting functions, and a rule of thumb states that a good fit is achieved
when $\chi^2/\nu \sim O(1)$. In Fig.~\ref{fig:co}(d), we find that including the
asymptotic correlation term, $c$, indeed describes the behavior of $C(l)$
better, since it makes the reduced $\chi^2$ statistic maintain its value
around $O(1)$ throughout the checked range of $p$, while the pure exponential
function without $c$ becomes a poor description for the same data at $p
\gtrsim 0.54$.
One therefore finds that the threshold value determined
numerically is consistent with the analytic prediction given above, although
the result is subject to a larger uncertainty than the case of $p_{c1}$.

\section{Summary}
\label{sec:sum}

In summary, we set upper bounds for $p_{c1}$ and $p_{c2}$ of the
EBT structure; that is, $p_{c1} < p^{\ast}_3 = 0.355059$ and
$p_{c2} \le 1/2$. In addition, we obtained $p_{c2} \ge 1/2$ as well,
which confirms the argument in \cite{future} that $p_{c2}=1/2$.
Since the upper threshold has been particularly under
debate, we tried to settle the issue to a large extent by showing that the
RG method developed for hierarchical structures led to the bounds of $p_{c2}$.
We also demonstrated that the correlation analysis yielded a consistent
result with this RG argument. Both of these analytical and numerical
approaches disprove the duality conjecture that the EBT and its dual lattice
are related in such a way that the upper threshold for the EBT, $p_{c2}$,
and the lower threshold for the dual lattice, $\bar{p}_{c1}$, sum up to
$p_{c2}+\bar{p}_{c1}=1$~\cite{nogawa}, since the lower threshold for the dual
lattice is numerically determined as $\bar{p}_{c1}=0.436(1)$ to good
precision~\cite{nogawa}. The possible value of $p_{c2}$ given in this work
clearly shows that $\bar{p}_{c1} +p_{c2}<1$, as already reported for other
hyperbolic lattices~\cite{baek}. This observation confirms that such a
duality relation requires transitivity~\cite{ben}, which does not hold for
the EBT.

From a methodological viewpoint, our analysis in this work relies
largely upon the regularity of the underlying structure. There also
exist other regular hierarchical structures such as the Apollonian
networks~\cite{apollo,apollo2} and flower networks~\cite{rozen}, where the
percolation problem has been studied by means of recursion.
The extension and usefulness of our subgraph analysis for these cases remain
to be investigated.

\section*{Acknowledgement}
We are grateful to Dr. Sebastian Bernhardsson for discussions.
We acknowledge the support from the Swedish Research Council
with Grant No. 621-2002-4135.
This research was conducted using the resources of High Performance
Computing Center North (HPC2N).


\begin{thebibliography}{17}
\expandafter\ifx\csname natexlab\endcsname\relax\def\natexlab#1{#1}\fi
\providecommand{\bibinfo}[2]{#2}
\ifx\xfnm\relax \def\xfnm[#1]{\unskip,\space#1}\fi
\bibitem[{Broadbent and Hammersley(1957)}]{broad}
\bibinfo{author}{S.~R. Broadbent}, \bibinfo{author}{J.~M. Hammersley},
\newblock \bibinfo{title}{Percolation processes: I. crystals and mazes},
\newblock \bibinfo{journal}{Proc. Cambridge Philos. Soc.} \bibinfo{volume}{53}
  (\bibinfo{year}{1957}) \bibinfo{pages}{629}.
\bibitem[{Stauffer and Aharony(2003)}]{stauffer}
\bibinfo{author}{D.~Stauffer}, \bibinfo{author}{A.~Aharony},
  \bibinfo{title}{Introduction to Percolation Theory},
  \bibinfo{publisher}{Taylor \& Francis}, \bibinfo{address}{London},
  \bibinfo{edition}{2} edition, \bibinfo{year}{2003}.
\bibitem[{Baek et~al.(2009)Baek, Minnhagen, and Kim}]{baek}
\bibinfo{author}{S.~K. Baek}, \bibinfo{author}{P.~Minnhagen},
  \bibinfo{author}{B.~J. Kim},
\newblock \bibinfo{title}{Percolation on hyperbolic lattices},
\newblock \bibinfo{journal}{Phys. Rev. E} \bibinfo{volume}{79}
  (\bibinfo{year}{2009}) \bibinfo{pages}{011124}.
\bibitem[{Nogawa and Hasegawa(2009)}]{nogawa}
\bibinfo{author}{T.~Nogawa}, \bibinfo{author}{T.~Hasegawa},
\newblock \bibinfo{title}{Monte Carlo simulation study of the two-stage
  percolation transition in enhanced binary trees},
\newblock \bibinfo{journal}{J. Phys. A} \bibinfo{volume}{42}
  (\bibinfo{year}{2009}) \bibinfo{pages}{145001}.
\bibitem[{Gilbert et~al.(2010)Gilbert, Schaack, II, Brindley, and
  Feschotte}]{gilbert}
\bibinfo{author}{C.~Gilbert}, \bibinfo{author}{S.~Schaack},
  \bibinfo{author}{J.~K.~Pace~II}, \bibinfo{author}{P.~J. Brindley},
  \bibinfo{author}{C.~Feschotte},
\newblock \bibinfo{title}{A role for host–parasite interactions in the
  horizontal transfer of transposons across phyla},
\newblock \bibinfo{journal}{Nature} \bibinfo{volume}{464}
  (\bibinfo{year}{2010}) \bibinfo{pages}{1347}.
\bibitem[{Baek et~al.(2009)Baek, Minnhagen, and Kim}]{com}
\bibinfo{author}{S.~K. Baek}, \bibinfo{author}{P.~Minnhagen},
  \bibinfo{author}{B.~J. Kim},
\newblock \bibinfo{title}{Comment on `Monte Carlo simulation study of the
  two-stage percolation transition in enhanced binary trees'},
\newblock \bibinfo{journal}{J. Phys. A} \bibinfo{volume}{42}
  (\bibinfo{year}{2009}) \bibinfo{pages}{478001}.
\bibitem[{Nogawa and Hasegawa(2009)}]{reply}
\bibinfo{author}{T.~Nogawa}, \bibinfo{author}{T.~Hasegawa},
\newblock \bibinfo{title}{Reply to the comment on `Monte Carlo simulation study
  of the two-stage percolation transition in enhanced binary trees'},
\newblock \bibinfo{journal}{J. Phys. A} \bibinfo{volume}{42}
  (\bibinfo{year}{2009}) \bibinfo{pages}{478002}.
\bibitem[{Minnhagen and Baek(2010)}]{future}
\bibinfo{author}{P.~Minnhagen}, \bibinfo{author}{S.~K. Baek},
\newblock \bibinfo{title}{Analytic results for the percolation transitions of
  the enhanced binary tree},
\newblock \bibinfo{journal}{Phys. Rev. E} \bibinfo{volume}{82}
  (\bibinfo{year}{2010}) \bibinfo{pages}{011113}.
\bibitem[{Scullard(2006)}]{scu}
\bibinfo{author}{C.~R. Scullard},
\newblock \bibinfo{title}{Exact site percolation thresholds using a
  site-to-bond transformation and the star-triangle transformation},
\newblock \bibinfo{journal}{Phys. Rev. E} \bibinfo{volume}{73}
  (\bibinfo{year}{2006}) \bibinfo{pages}{016107}.
\bibitem[{Ziff(2006)}]{ziff06}
\bibinfo{author}{R.~M. Ziff},
\newblock \bibinfo{title}{Generalized cell-dual-cell transformation and exact
  thresholds for percolation},
\newblock \bibinfo{journal}{Phys. Rev. E} \bibinfo{volume}{73}
  (\bibinfo{year}{2006}) \bibinfo{pages}{016134}.
\bibitem[{Ziff and Gu(2009)}]{ziff09}
\bibinfo{author}{R.~M. Ziff}, \bibinfo{author}{H.~Gu},
\newblock \bibinfo{title}{Universal condition for critical percolation
  thresholds of kagom\'{e}-like lattices},
\newblock \bibinfo{journal}{Phys. Rev. E} \bibinfo{volume}{79}
  (\bibinfo{year}{2009}) \bibinfo{pages}{020102(R)}.
\bibitem[{Vuorio(1974)}]{vuo}
\bibinfo{author}{M.~Vuorio},
\newblock \bibinfo{title}{A method to estimate the critical probability in bond
  percolation problems},
\newblock \bibinfo{journal}{J. Chem. Phys.} \bibinfo{volume}{60}
  (\bibinfo{year}{1974}) \bibinfo{pages}{846}.
\bibitem[{Boettcher et~al.(2009)Boettcher, Cook, and Ziff}]{boet}
\bibinfo{author}{S.~Boettcher}, \bibinfo{author}{J.~L. Cook},
  \bibinfo{author}{R.~M. Ziff},
\newblock \bibinfo{title}{Patch percolation on a hierarchical network with
  small-world bonds},
\newblock \bibinfo{journal}{Phys. Rev. E} \bibinfo{volume}{80}
  (\bibinfo{year}{2009}) \bibinfo{pages}{041115}.
\bibitem[{Benjamini and Schramm(2000)}]{ben}
\bibinfo{author}{I.~Benjamini}, \bibinfo{author}{O.~Schramm},
\newblock \bibinfo{title}{Percolation in the hyperbolic plane},
\newblock \bibinfo{journal}{J. Am. Math. Soc} \bibinfo{volume}{14}
  (\bibinfo{year}{2000}) \bibinfo{pages}{487--507}.
\bibitem[{J.~S.~Andrade et~al.(2005)J.~S.~Andrade, Herrmann, Andrade, and
  da~Silva}]{apollo}
\bibinfo{author}{J.~S.~Andrade, Jr.}, \bibinfo{author}{H.~J. Herrmann},
  \bibinfo{author}{R.~F.~S. Andrade}, \bibinfo{author}{L.~R. da~Silva},
\newblock \bibinfo{title}{Apollonian networks: Simultaneously scale-free, small
  world, Euclidean, space filling, and with matching graphs},
\newblock \bibinfo{journal}{Phys. Rev. Lett.} \bibinfo{volume}{94}
  (\bibinfo{year}{2005}) \bibinfo{pages}{018702}.
\bibitem[{Auto et~al.(2008)Auto, Moreira, Herrmann, and
  J.~S.~Andrade}]{apollo2}
\bibinfo{author}{D.~M. Auto}, \bibinfo{author}{A.~A. Moreira},
  \bibinfo{author}{H.~J. Herrmann}, \bibinfo{author}{J.~S.~Andrade, Jr.},
\newblock \bibinfo{title}{Finite-size effects for percolation on Apollonian
  networks},
\newblock \bibinfo{journal}{Phys. Rev. E} \bibinfo{volume}{78}
  (\bibinfo{year}{2008}) \bibinfo{pages}{066112}.
\bibitem[{Rozenfeld and ben Avraham(2007)}]{rozen}
\bibinfo{author}{H.~D. Rozenfeld}, \bibinfo{author}{D.~ben-Avraham},
\newblock \bibinfo{title}{Percolation in hierarchical scale-free nets},
\newblock \bibinfo{journal}{Phys. Rev. E} \bibinfo{volume}{75}
  (\bibinfo{year}{2007}) \bibinfo{pages}{061102}.

\end{thebibliography}

\end{document}